\documentclass{aa}
\usepackage{graphicx}

\begin{document}

\title{Supernova 1998bw - The final phases
       \thanks{Based on observations collected at the European
               Southern Observatory, La Silla and Paranal, Chile and
               on observations made with the NASA/ESA \emph{Hubble
               Space Telescope}, obtained at the Space Telescope
               Science Institute, which is operated by the Association
               of Universities for Research in Astronomy, Inc., under
               NASA contract NAS 5-26555.}}

\author{J.~Sollerman,\inst{1,2}
        S.~T.~Holland,\inst{3}
        P.~Challis,\inst{4}
        C.~Fransson,\inst{2}
        P.~Garnavich,\inst{3}
        R.~P.~Kirshner,\inst{4}
        C.~Kozma,\inst{2}
        B.~Leibundgut,\inst{1}
        P.~Lundqvist,\inst{2}
        F.~Patat,\inst{1}
        A.~V.~Filippenko,\inst{5}
        N.~Panagia,\inst{6}
        J.~C.~Wheeler\inst{7}
}

\offprints{Jesper Sollerman, \email{jesper@astro.su.se}}

\institute{European Southern Observatory,
           Karl-Schwarzschild-Strasse 2, 
           D--85748 Garching bei M\"unchen, Germany 
           \and
           Stockholm Observatory,
           Department of Astronomy, SCFAB, 
           SE--106 91 Stockholm, Sweden
           \and
           Physics Department,
           University of Notre Dame,
           Notre Dame, IN, US
           \and
           Harvard--Smithsonian Center for Astrophysics,
           60 Garden Street, Cambridge, US
           \and
           Department of Astronomy,
           University of California,
           Berkeley, CA, US
           \and
           Space Telescope Science Institute,
           3700 San Martin Drive,
           Baltimore, MD, US
           \and
           Department of Astronomy,
           University of Texas,
           Austin, TX, US
}

\date{Received / Accepted}

\authorrunning{Jesper Sollerman et~al.}

\titlerunning{SN 1998bw, The final phases.}

\abstract{%

     The probable association with \object{GRB~980425} immediately put
\object{SN~1998bw} at the forefront of supernova research.  Here, we
present revised late-time $BVRI$ light curves of the supernova, based
on template images taken at the VLT\@.  To follow the supernova to the
very last observable phases we have used \emph{HST}/STIS\@. Deep
images taken in June and November 2000 are compared to images taken in
August 2001.  The identification of the supernova is firmly
established. This allows us to measure the light curve to $\sim1000$
days past explosion.  The main features are a rapid decline up to more
than 500 days after explosion, with no sign of complete positron
trapping from the \element[][56]{Co} decay. Thereafter, the light
curve flattens out significantly.  One possible explanation is
powering by more long lived radioactive isotopes, if they are
abundantly formed in this energetic supernova.


}

\maketitle

\section{Introduction}

     Supernova (SN) 1998bw was discovered in April 1998 in a spiral
arm of the face-on SB galaxy \object{ESO~184-G82} at a redshift of
$z=0.0085$ (Galama et~al.\ 1998a).  This exceptional type Ic supernova
was born famous because of its positional and temporal coincidence
with the gamma-ray burst \object{GRB~980425} (Galama et~al.\ 1998b).
This association was further supported by the very early onset of the
bright radio luminosity (Kulkarni et~al.\ 1998). Also in the optical,
the supernova was unusually luminous and showed very high expansion
velocities (Patat et~al.\ 2001).

     The earliest photometric evolution of \object{SN~1998bw} was
presented in Galama et~al.\ (1998b). McKenzie \& Schaefer (1999)
monitored the early exponential decay, and the late phases were added
by Sollerman et~al.\ (2000). Patat et~al.\ (2001) combined these data
to make a rough bolometric light curve, which has been the basis of
several model investigations (Nomoto et~al.\ 2001; Nakamura et~al.\
2001a).  The aim of this paper is to improve and extend this
light curve to the final phases of \object{SN~1998bw}.

     Modelling of the early light curve and spectra suggested an
extremely energetic explosion [$(2-5)\times~10^{52}$~erg] of a massive
star, composed mainly of carbon and oxygen (Iwamoto et~al.\ 1998;
Woosley et al. 1999).  As this kinetic energy is more
than ten times that of a canonical core-collapse supernova, the term
hypernova was suggested for this event.  The mass of
\element[][56]{Ni} needed to power the early light curve in these
early models, 0.5--0.7 ${~\rm M}_\odot$, is much larger than the $\la 0.1$
${~\rm M}_\odot$
typical for ``normal'' core-collapse SNe (e.g., Sollerman
2002).

     This large mass of \element[][56]{Ni} was disputed by H\"oflich
et al. (1999), who suggested that the early light curve
could be reproduced by a normal SN~Ic, given the right viewing angle
and degree of asymmetry.  The modelling of Sollerman et~al.\ (2000)
indicated, however, that at least 0.3 ${~\rm M}_\odot$ of \element[][56]{Ni}
was required to power the late emission of the supernova. This
conclusion was later supported by Nakamura et~al.\ (2001a), see also
Mazzali et~al.\ (2001).

     In this paper we focus on the very late light curve of
\object{SN~1998bw}.  In Sect.~2 we present new VLT observations and the
technique used to achieve the revised optical light curves.  Here we
also present some late near-IR photometry.  In Sect.~3 we present the very
late \emph{Hubble Space Telescope} (\emph{HST}) observations of
\object{SN~1998bw}.  The results are presented in Sect.~4.  After a
discussion in Sect.~5, we briefly summarize our conclusions in Sect.~6.


\section{Ground Based Observations}

\subsection{Late-Time Optical Photometry}

     The photometry discussed here is based on the late observations
presented in Sollerman et~al.\ (2000; hereafter S00). S00 obtained
data at eight epochs between 32 and 503 days past explosion, using the
ESO 3.6m telescope on La Silla and the Very Large Telescope (VLT) on
Paranal. Here we assume the date of explosion to be 980425.9, the
moment of detection of \object{GRB~980425} (Soffitta et~al.\ 1998).

     To the S00 data set we have added the late time observations
presented in Patat et al.\ (2001; hereafter P01). These data were
primarily obtained using the ESO 3.6m telescope and are added in
particular to include the $B$-band photometry. We have also added
previously unpublished photometry obtained at the VLT\@.  The details
of the ground-based optical observations are shown in
Table~\ref{TABLE:logground}.


\begin{table*}[ht]
\begin{center}
\caption[]{Log of ground based observations}
\begin{tabular}{lccccc}
\hline
\noalign{\smallskip}

Date & MJD$^{a}$ & Epoch$^{b}$ & Telescope$^{c}$ & Bands & Source \\

\noalign{\smallskip}
\hline
\noalign{\smallskip}

13 Sep.\ 98   & 1069 & 140 & 3.6m & $V\!RI$  & S00 \\
26 Nov.\ 98   & 1143 & 214 & 3.6m & $V\!RI$  & S00 \\
18 Mar.\ 99   & 1255 & 326 & 3.6m & $V\!RI$  & S00 \\
18 Mar.\ 99   & 1255 & 326 & 0.9m & $B$    & P01 \\    
08 Apr.\ 99   & 1276 & 347 & 3.6m & $BV\!R$  & P01 \\
12 Apr.\ 99   & 1280 & 351 & 3.6m & $I$    & P01 \\
17 Apr.\ 99   & 1285 & 356 & VLT  & $V\!RI$  & S00 \\
17 Apr.\ 99   & 1285 & 356 & VLT  & $B$    & This work \\
21 May 99     & 1319 & 390 & 3.6m & $BV\!RI$ & P01 \\
13 Jun.\ 99   & 1342 & 413 & VLT  & $V\!RI$  & S00 \\
17 Jun.\ 99   & 1346 & 417 & 3.6m & $BV\!RI$ & P01 \\
09 Aug.\ 99   & 1399 & 470 & 3.6m & $V\!RI$  & S00 \\
11 Sep.\ 99   & 1432 & 503 & VLT  & $V\!RI$  & S00 \\
11 Sep.\ 99   & 1432 & 503 & VLT  & $B$    & This work \\
16 Oct.\ 99   & 1467 & 538 & VLT  & $BV\!RI$ & This work \\
01 Apr.\ 00   & 1635 & 706 & VLT  & $BV\!RI$ & This work \\
Oct./Nov.\ 00 & $\sim1836$ & $\sim907$  & VLT & $BV\!RI$ & This work \\

\noalign{\smallskip}
\hline
\end{tabular}
\label{TABLE:logground}
\end{center}
$^{a}$ {MJD$=$Julian Date$-$240000.5}\\
$^{b}$ Epoch in days past 980425.9\\
$^{c}$ VLT=UT1+FORS1, 3.6m=ESO-3.6m+EFOSC2,\\
$^{}\;\,$ 0.9m=ESO-Dutch-0.92m\\
\end{table*}

\begin{figure*} 
\includegraphics{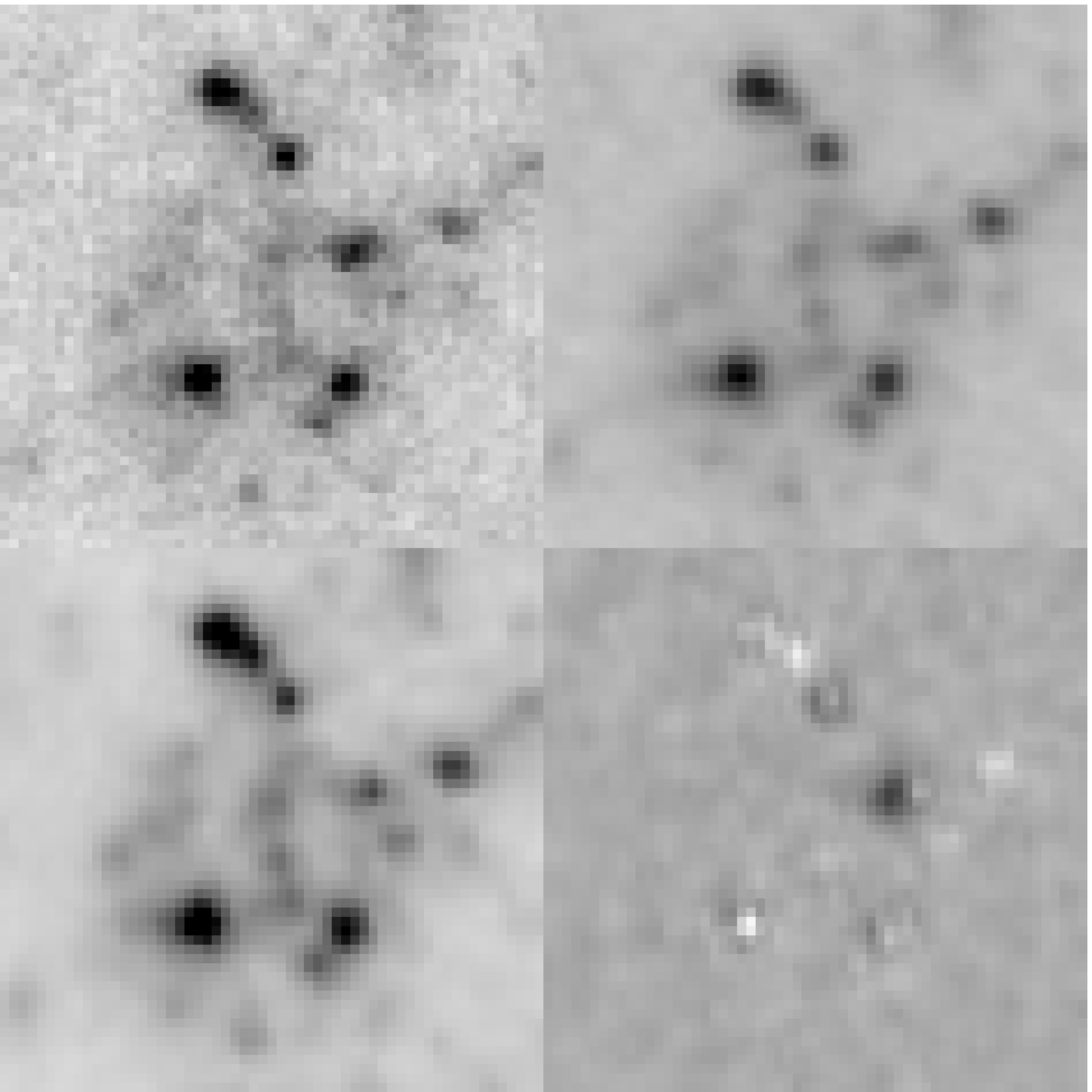}
\caption{\emph{HST}/STIS CLEAR images of \protect\object{SN~1998bw}
and its environment.  The images were obtained in 2000 June (upper
left), 2000 November (upper right), and 2001 August (lower left). The
field of view is merely $1\farcs5 \times 1\farcs5$. Lower right shows
a subtraction between the first and the final frame. The supernova is
clearly identified.  The apparent differences in resolution are due to
differences in the dithering procedure, as explained in the text.}
\label{FIGURE:hst}
\end{figure*}

     As discussed in S00 and P01, the photometry of \object{SN~1998bw}
is fairly complicated. This is because the supernova is superposed on
an
\ion{H}{ii} region, and sits in a rather complex region of the host
galaxy.  This was strikingly revealed by the \emph{HST}/STIS image
published by Fynbo et~al.\ (2000), where several objects can be seen
within $1\arcsec$ of the position of the supernova.  This is here
shown in detail in Fig.~\ref{FIGURE:hst}.


     This means that even sophisticated point-spread function (PSF)
fitting photometry may not be adequate to accurately measure the SN
brightness at late phases. S00 used {\sc DAOPhot} to estimate the
errors due to background subtraction, while P01 used the {\sc SNOOPY}
software. Although these methods were consistent, the very complex
background of \object{SN~1998bw} as observed with \emph{HST} called
for a revision of the light curve.  We decided to try another method
to establish the late time light curve of this important supernova.

     We therefore obtained deep $B$, $V$, $R$, and $I$ imaging of the
region containing the supernova using FORS1 on the VLT\@. These
observations were obtained in service mode under excellent atmospheric
conditions in
early October and in early November 2000, some 900 days after the
supernova explosion.  In each filter band 6--9 frames
were coadded to achieve a deep combined image. The $B$ and $V$ frames
have a total exposure time of 90 minutes each, while the $R$ and $I$
frames have 120 minutes. All combined frames have good image quality,
with a FWHM less than one arcsecond.

     In these images, no sign of the supernova could be detected,
i.e., the background on the position of the supernova appeared
smooth. These frames were therefore used as template frames to
subtract the background from all the previous images (see e.g.,
Filippenko et~al.\ 1986 for an early account of this method).  The
detailed procedure used is outlined in Schmidt et~al.\ (1998), and we
have used the same software as used by the High-$z$ Supernova Search
Team for galaxy subtraction, magnitude measurements and error
estimates.

     After the usual bias subtraction and flat fielding, all images
were aligned with the template image in the same filter band.  The
image with the best seeing was then convolved to match the seeing of
the other image. Thereafter the image was scaled to the same intensity
as the template, and the template was subtracted from the image in a
region around the supernova.  The program {\sc DoPhot} (Schechter et al.
1993) was used to measure the magnitudes of the
supernova and of the local standard stars in the subtracted image.

     A PSF was constructed from stars in the field using {\sc DAOPhot}
(Stetson 1987). It was used to add ten artificial stars with the same
flux as the supernova to user-specified locations on the original
frame. The magnitudes of these stars were then retrieved using the
same technique as for the supernova. This method was used to assess
the accuracy of our measurements, and the standard deviations in the
differences in the retrieved magnitudes of the artificial stars were
adopted as the errors on the supernova magnitudes in
Table~\ref{TABLE:groundmags}.


\begin{table*}[ht]
\begin{center}
\caption{Supernova Magnitudes}
\begin{tabular}{lcccccccc}
\hline
\noalign{\smallskip}

Epoch & B & Berr & V & Verr & R & Rerr & I & Ierr \\

\noalign{\smallskip}
\hline
\noalign{\smallskip}

140 & --    & --    & 17.38 & 0.027 & 17.06 & 0.019 & 16.77 & 0.007 \\
214 & --    & --    & 18.76 & 0.025 & 18.14 & 0.032 & 17.95 & 0.051 \\
326 & 21.21 & 0.05 & 20.80 & 0.011 & 19.87 & 0.007 & 19.72 & 0.020 \\
347 & 21.56 & 0.011 & 21.15 & 0.025 & 20.21 & 0.018  &  --   &  --   \\
351 & --    & --    &   --  &  --   &  --   &  --   & 20.10 & 0.024 \\
356 & 21.66 & 0.069  & 21.30 & 0.037 & 20.27 & 0.017 & 20.03 & 0.025 \\
390 & 22.19 & 0.010 & 21.78 & 0.029 & 20.98 & 0.030 & 20.95 & 0.068  \\
413 & --    & --    & 22.12 & 0.029 & 21.18 & 0.016 & 21.03 & 0.029 \\
417 & 22.57 & 0.05  & 22.16 & 0.025 & 21.32 & 0.030 & 21.13 & 0.036  \\
470 & --    &  --   & 22.86 & 0.032 & 22.21 & 0.052 & 22.08 & 0.099 \\
503 & 23.53 & 0.043 & 23.07 & 0.067 & 22.59 & 0.041 & 22.50 & 0.103 \\
538 & --    & --    & 23.43 & 0.33  & 23.05 & 0.082 & 23.25 & 0.465 \\
706 & $>25.1$  &  --  & $>25.1$ & --  & $>25.1$ & -- & $>24.7$ & -- \\

\noalign{\smallskip}
\hline
\end{tabular}
\label{TABLE:groundmags}
\end{center}
\end{table*}


     In all cases, the success of the subtraction has been manually
checked. Indeed, very clean results were achieved, where only the
supernova remains on the subtracted frame. This is due to the fact
that many of these observations have been obtained with the same
telescope/instrument/CCD setup. Similar seeing conditions are also
important, and in particular a high quality template frame is crucial.

\begin{figure*}
\includegraphics{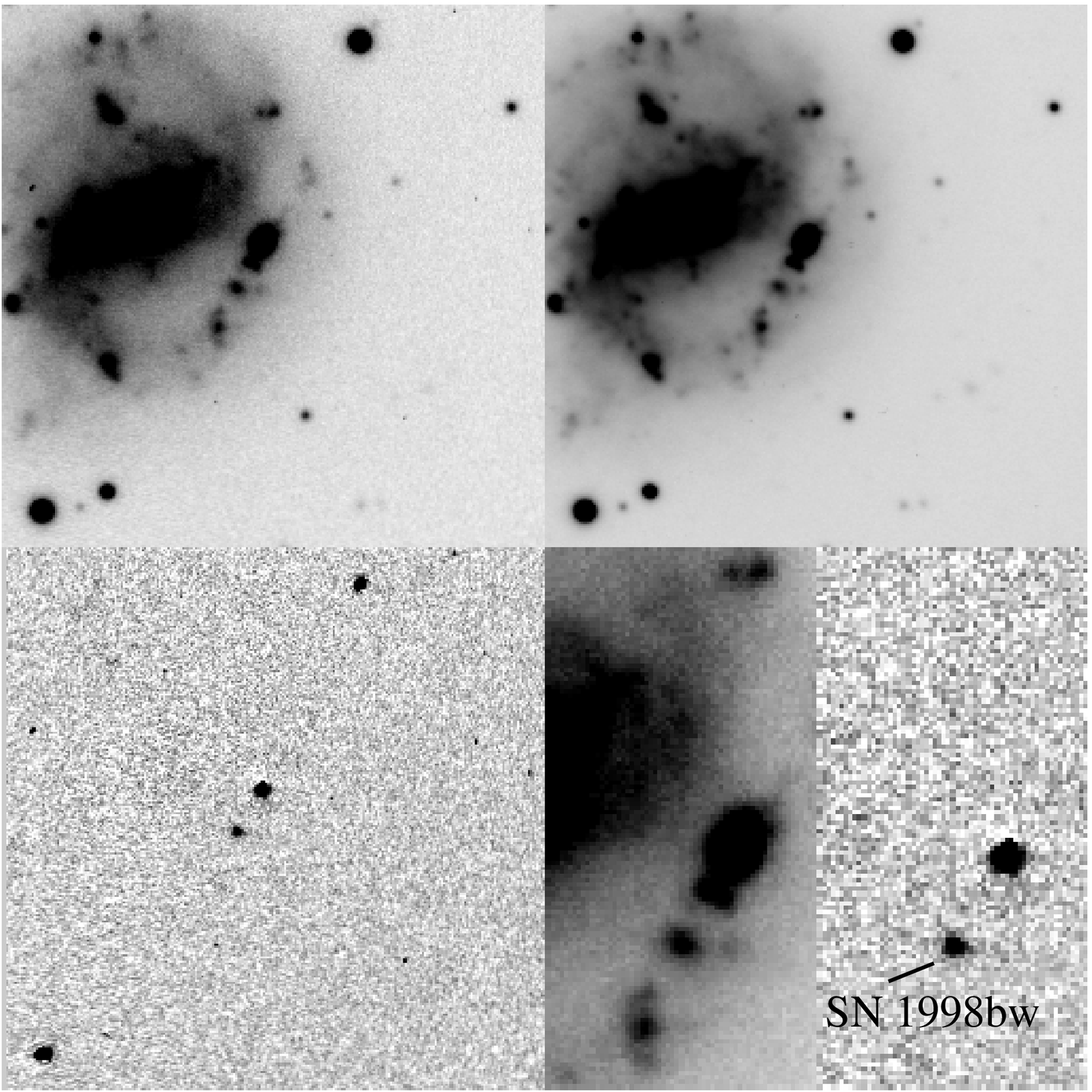}
\caption{Illustration of the template subtraction method.  Upper left
shows the supernova and its galaxy as observed with the VLT in 1999
September, 503 days past explosion. Upper right is the template
image. North is up and East to the left; the field of view is about
$52\arcsec\times52\arcsec$.  The subtraction is shown in the lower
left. Only the supernova, as well as some residuals from stars that
are saturated in the template frame, are seen.  Lower right shows a
blowup of the supernova region, before and after subtraction. Note how
the galaxy is completely subtracted.}
\label{FIGURE:subtraction}
\end{figure*}

     In Fig.~\ref{FIGURE:subtraction} we show an example of the region
of the supernova in our template frame, and in our frame from day
503. This was the last epoch used in S00 and P01. Also shown is the
subtraction, which clearly shows the supernova.


     The instrumental magnitudes were finally converted using local
standards measured by Galama et~al.\ (1998b). This worked well for the
$B$-band, where four standard stars showed a standard deviation of
$\le0.03$ magnitudes. Most of this scatter is in fact due to the
neglect of color terms, which, however, do not affect the supernova as
we chose a primary standard with similar color as the SN itself.  For
the other bands, all Galama standards were saturated in the long
exposures required at these late phases. We then established
secondary, faint local standards. The magnitudes of four of these
local standards are presented in Table~\ref{TABLE:stds}, together with
their angular offsets from the supernova.  The standard deviations for
these stars measured over all the epochs were never higher than 0.03
magnitudes. Adding this to the uncertainty reported by Galama et~al.\
for their standards, we consider the accuracy of the secondary
standards to be about 0.06 magnitudes.


\begin{table}[ht]
\begin{center}
\caption{Magnitudes of Faint Local Standards}
\begin{tabular}{lccc}
\hline
\noalign{\smallskip}

Offsets from SN & $V$ & $R$ & $I$ \\

\noalign{\smallskip}
\hline
\noalign{\smallskip}

 $46\arcsec$~N, $17\arcsec$~W & 19.77 & 19.03 & 18.45\\
 $50\arcsec$~N, $27\arcsec$~W & 21.13 & 19.95 & 18.69\\
 $17\arcsec$~N, $26\arcsec$~W & 21.64 & 20.44 & 19.12\\
 $30\arcsec$~S, $17\arcsec$~E & 21.24 & 20.37 & 19.63\\

\noalign{\smallskip}
\hline
\end{tabular}
\label{TABLE:stds}
\end{center}
\end{table}


\subsection{Late Upper Limits}

     At our last epoch of VLT observations prior to the template
images, i.e., at 706 days past explosion, the supernova is not
detected in the subtracted images.  We have used these difference
images to estimate upper limits on the supernova magnitude in the
following way.  We measured the fluxes in a number of apertures
located next to the supernova. From these measurements we estimated
the standard deviations in a 5 pixel radius aperture, corresponding to
the image FWHM\@. This estimate was then used to acquire $3\sigma$
upper limits within the aperture.  These limits are 25.1, 25.1, 25.1,
and 24.7 in $B$, $V$, $R$, and $I$, respectively. This procedure is
more conservative than a simple measurement of the statistics in the
region of the supernova. We also verified that a $R=25.1$ supernova
would indeed have been detected by adding such a star to the original
frame, and recovering the magnitude in the subtracted frame. This
would thus provide conservative upper limits on the flux of the
supernova at this epoch. The main assumption is that the supernova was
indeed absent in the template frame.  This is further elaborated
below.

\subsection{The Near-Infrared}

     Here, we present the last infrared (IR) observations of
\object{SN~1998bw} of which we are aware.

     \object{SN~1998bw} was observed using ISAAC on the VLT on 1999
May 1, 370 days after the explosion.  The observations were obtained
in Autojittermode, offsetting the telescope within a $60\arcsec$ wide
box. Five frames were obtained in the $J$-band (1.11--1.39 $\mu$m),
each one composed of a stack of five 30 second exposures. The total
exposure time in the $J$-band was thus 750 seconds.

     In the $H$-band (1.50--1.80 $\mu$m) five frames composed of five
20 seconds exposures yielded a total exposure time of 500 seconds.  In
addition, an infrared spectrum was obtained at this epoch, but the
signal-to-noise ratio of this spectrum is low, and we will not discuss
it further. Instead, we will use the two acquisition images to
determine the IR contribution at these late phases.

     The reductions were done using IRAF\footnote{Image Reduction and
Analysis Facility (IRAF), a software system distributed by the
National Optical Astronomy Observatories (NOAO).}.  From each image we
subtracted a mean sky, constructed from combining all the other frames
with suitable clipping to remove objects. Flat field corrections were
applied using dome flat-fields.  The images were then stacked and
combined. The FWHM in the combined frames were $0\farcs4$ and
$0\farcs55$ for $J$ and $H$, respectively.

     For the calibration of the photometry, we used the zero points
determined for this specific night at the VLT, from observations of
photometric standards. To check the stability of the zero point, we
also re-measured three local standard stars from the three earlier
epochs obtained with NTT/SOFI (P01). The standard deviation in the
standard stars on the three SOFI nights were 0.04 and 0.02 magnitudes
in $J$ and $H$, respectively. The magnitudes agreed with the ISAAC
measurements to better than 0.05 magnitudes, with a scatter of 0.02
mag.

     In the combined frames, the SN magnitude were measured with PSF
photometry using {\sc DAOPhot} within IRAF\@.  The measured magnitudes
are $J=20.16$ and $H=19.54$.  The aperture correction was obtained
from the three local standards, with a negligible scatter. The {\sc
DAOPhot} measurement error was 0.03 ($J$) and 0.06 ($H$) magnitudes.
The total error budget for the photometry is therefore certainly
better than 0.1 magnitudes in both bands.  These results should be
treated with some caution, however, as we know from the optical data
that PSF fitting may not be the best technique for this supernova.

\section{The Last Phases with HST}

\subsection{STIS Observations}

     In order to follow the supernova to very late times, and to get a
better understanding of the supernova environment, we have obtained
detailed imaging at four epochs using the Space Telescope Imaging
Spectrograph (STIS) onboard the \emph{HST}.  A log of our HST
observations is given in Table~\ref{TABLE:hstlog}.


\begin{table}
\begin{center}
\caption{Log of HST Observations}
\begin{tabular}{lcccc}

\hline
\noalign{\smallskip}

Date & MJD & Epoch & Exp. Time & Exp. Time \\
 (UT)  &    &  (days)    & CL (s) & LP (s) \\

\noalign{\smallskip}
\hline
\noalign{\smallskip}

11 June 00 & 1707 & 778 & 1240 & 1185 \\
25 June 00 & 1721 & 792 & 5672 & --- \\
21 Nov. 00 & 1869 & 940 & 4000 & 1319 \\
28 Aug. 01 & 2150 & 1221 & 5674 & --- \\

\noalign{\smallskip}
\hline

\end{tabular}
\label{TABLE:hstlog}
\end{center}
\end{table}



     The observations on 2000 June 25, 2000 Nov.\ 21, and 2001 Aug.\
28 were taken as part of the Supernova Intensive Study programme
(P.I.\ Kirshner).  The data from 2000 June 11 was taken as part of the
Survey of Host Galaxies of Gamma-Ray Bursts (P.I.\ Holland, Holland
et~al.\ 2000).  The total exposure times in the 50CCD (clear, CL)
aperture were 1240, 5672, 4000, and 5674 seconds on the four epochs,
respectively.
The 50CCD aperture has a central wavelength of 5850~\AA~and a width of
4410~\AA.

     In addition to the CL frames we also obtained some images in the
F20X50LP (long pass, LP) filter at two epochs. The total exposure
times were 1185 and 1319 seconds, respectively.
The F28x50LP aperture has a central wavelength of 7320~\AA~and a width
of 2720~\AA.

     The 2000 June 11 images were taken using a four-point
STIS-CCD-BOX dithering pattern with shifts of 2.5 pixels
($=0\farcs127$) between exposures.  The other images were dithered
using the same pattern with shifts of 5 pixels ($= 0\farcs254$).

     The data were preprocessed through the standard STIS pipeline and
combined using the {\sc dither} (v2.0) software (Fruchter \& Hook
2002) as implemented in IRAF\@.  We selected a final output scale of
$0\farcs0254$/pixel, corresponding to exactly half the original size,
and set the ``pixfrac'' parameter to 0.6.  Since there were only two
images taken using the LP filter on 2000 Nov.\ 21 we had to increase
``pixfrac'' to 1.0 for these data in order to fill all pixels in the
output frame.  This resulted in a reduction in the image quality of
this data compared to the drizzled 2000 June 11 LP image.  The six
drizzled images were finally reregistered onto a common system.

\subsection{Photometry}


     The supernova is located in a complex region of the host galaxy
(Fig.~\ref{FIGURE:hst}).  There are several objects within a radius of
$1\arcsec$ from the supernova position, and \object{SN~1998bw} itself
is located on top of an extended, filamentary structure.  In order to
perform photometry of the supernova without contamination from the
underlying structure we again used the template subtraction method to
subtract the 2001 Aug.\ 28 CL image from each of the other CL images.
An example subtraction is given in Fig.~\ref{FIGURE:hst}.

     We then constructed PSFs for each subtracted CL image using {\sc
DAOPhot II}/{\sc AllStar} (Stetson 1987; Stetson \& Harris 1988) and
fit these to the image of the supernova on each subtracted frame.  We
computed aperture corrections for each frame using several isolated
stars.  The PSF magnitudes were corrected to an aperture with a radius
of $1\farcs108$, which Table 14.3 of the STIS Instrument Handbook
suggests contains 100\% of the flux from a point source.  These
corrections were about 0.1 magnitudes at all epochs.  The instrumental
magnitudes were converted to the AB system using the zero points of
Gardner et~al.\ (2000).

     In order to understand any systematic errors we performed a
series of artificial star tests on each CL image.  We added a set of
stars with known magnitudes comparable to that of the supernova to
locations on each frame where the underlying structure was similar to
the underlying structure of \object{SN~1998bw}.  We then subtracted
the template image and performed photometry on these artificial stars
in exactly the same way as for the supernova.  The resulting median
differences between the input and recovered magnitudes of the
artificial stars were always below 0.2 magnitudes. We regard this as a
conservative error on the difference photometry.

     No template image was taken using the LP filter. Therefore, no
reliable magnitudes could be obtained in this band. For the 2000 June
11 image we estimated the supernova magnitude by pure PSF fitting on
the drizzled image.  The estimated magnitude is
LP$_{\mathrm{PSF}}=25.02\pm0.07$.  Note that there was an error in the
LP aperture corrections used by Fynbo et~al.\ (2000).  That paper also
contains some color information on the surrounding objects, although a
correction of -0.51 mag should be added to all those LP magnitudes.
The PSF magnitude for the supernova will clearly also include some of
the light from the underlying background, which will result in
overestimating the flux from the supernova.  As mentioned above, the
2000 Nov.\ 21 LP frame is less useful. Visual inspection of that frame
does show the supernova, just as in the CL frame from the same epoch,
but we were not able to perform a reliable PSF estimate of its
magnitude, since the remaining PSF residuals were always large.
Therefore, only the CL magnitudes will be discussed below.


\section{Results}

\subsection{The Optical Light Curve from the Ground}

\begin{figure*}
\includegraphics{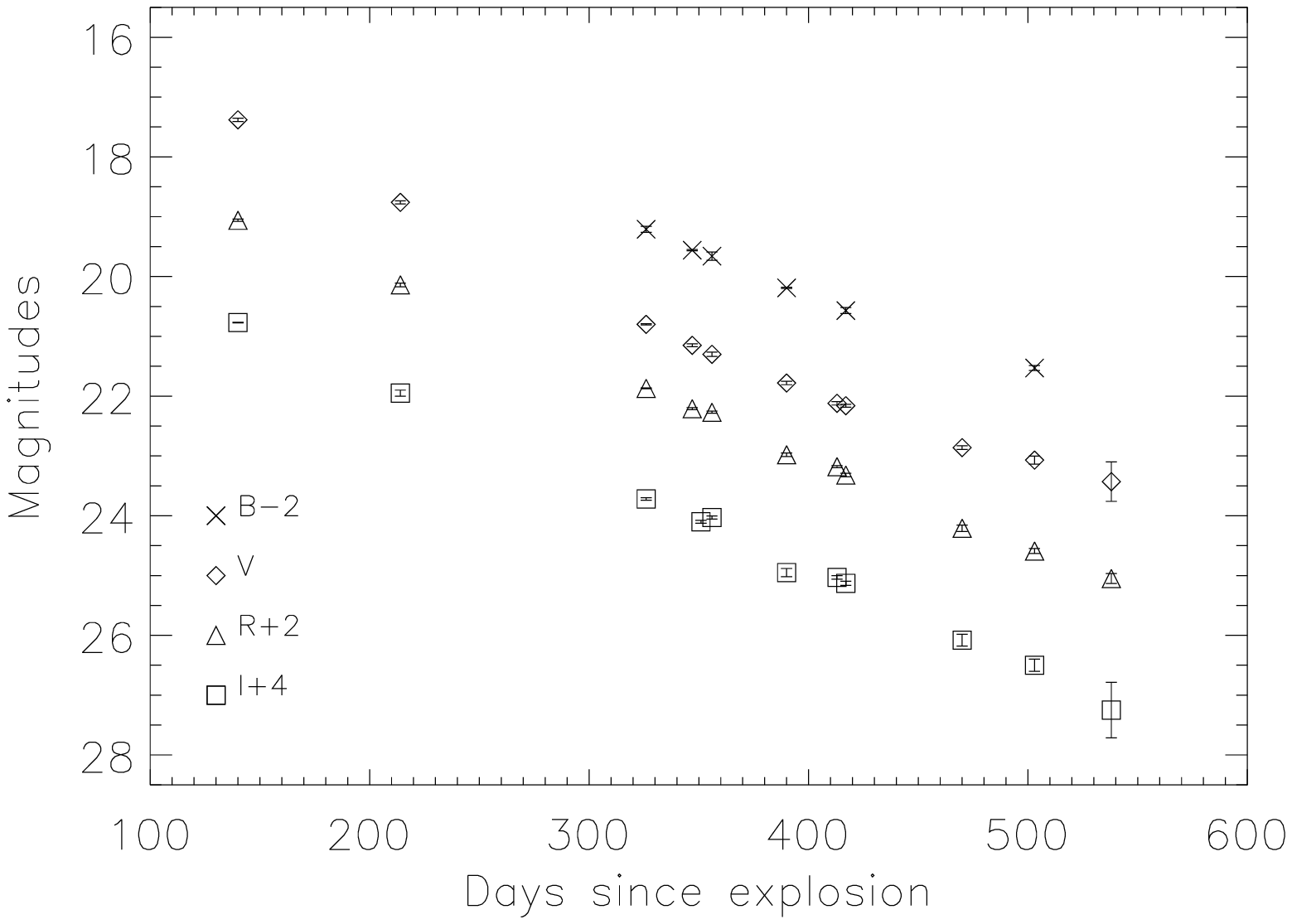}
\caption{Ground-based $B$, $V$, $R$, and $I$ photometry of
\protect\object{SN~1998bw} as obtained with the template subtraction
method.  The filter band light curves have been offset as indicated
for clarity.}
\label{FIGURE:lc}
\end{figure*}

     The results of the ground based photometry are presented in
Table~\ref{TABLE:groundmags}, and shown in Fig.~\ref{FIGURE:lc}.


     The light curves presented in Fig.~\ref{FIGURE:lc} are all very
smooth, with the exponential decay continuing all the way to the very
last data points, some 540 days past explosion.  The $R$ and $I$
curves, for example, are perfectly well fit by a linear decline of
1.5--1.6 magnitudes per 100 days.  This is in contrast to the results
published in S00 and P01, where the very late light curve appeared to
flatten out. We interpret the difference as due to the complex
background of \object{SN~1998bw}, and regard the new magnitudes
achieved with template subtraction as more accurate. Having seen the
very complex environment of the supernova in \emph{HST} detail this
may not be surprising. In fact, the total magnitude of the complex
region surrounding the supernova amounts to $m_{\mathrm{V}} \sim 21$.
In measuring supernovae some 2 magnitudes fainter than the background,
even small errors in the PSF fitting will amount to substantial errors
in the supernova magnitudes, an effect that will mimic a flattening of
the late light curve. Core-collapse supernovae are known to explode in
regions of star formation, and it is not unlikely that photometry of
other supernovae at very late phases is hampered by similar effects
(see Clocchiatti et~al.\ 2001 for a similar discussion).

     To investigate the total light emitted 
should try to build a bolometric light curve.  We begin by
constructing the $L_{BVRI}$ light curve, containing the optical
emission from 4400 to 7900 {\AA}. We have done this in the following
way.

\begin{figure*}
\includegraphics{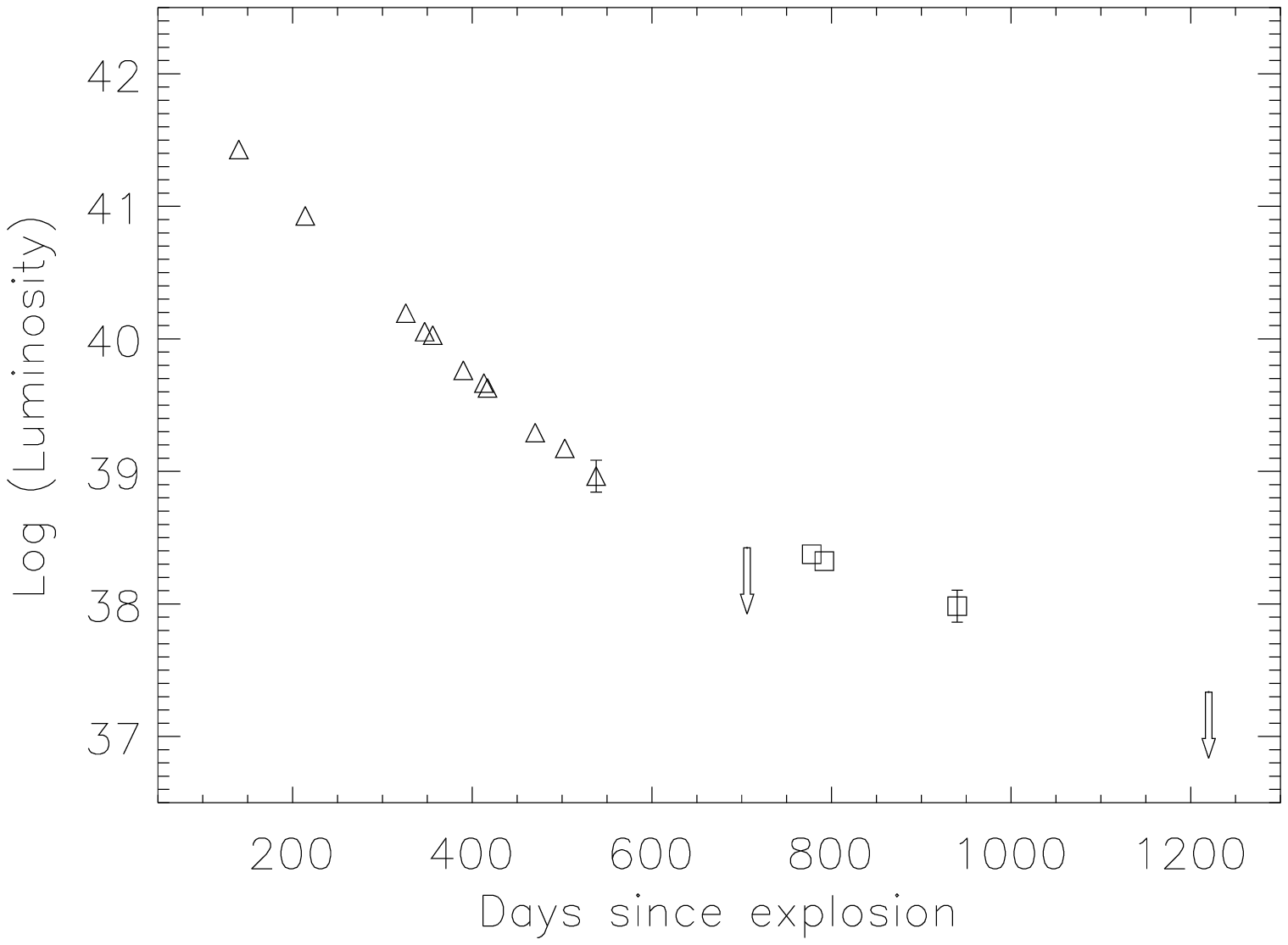}
\caption{The OIR light curve of \protect\object{SN~1998bw}. The
luminosities have been derived using the techniques described in the
text. A constant IR contribution of 42\%, as observed on day 370, has
been added to all phases.  The data up to 700 days are from ground
based observations, and the later epochs are obtained with
\emph{HST}. The two error bars plotted indicate the uncertainty from
0.3 mag errors in the photometry. The arrows represent $3\sigma$ upper
limits.}
\label{FIGURE:bolum}
\end{figure*}

     All magnitudes were converted to a monochromatic flux using the
conversions in Bessell (1979). To add the missing $B$-magnitudes, a
linear fit to these magnitudes was used, where the day 141 magnitude
from McKenzie \& Schaefer (1999) was included to constrain the early
evolution. The light curve slope in the $B$-band is then the same as
in the other bands.  The fluxes were corrected for an extinction of
$E(B-V)$=0.06 mag (Schlegel et~al.\ 1998; P01), and then the
monochromatic fluxes were simply integrated
from 4400 to 7900 {\AA}.  Finally we adopted a distance of 35 Mpc to
convert to luminosity ($z=0.0085$, $H_0=73$ km~s$^{-1}$~Mpc$^{-1}$).
To the resulting $L_{BVRI}$ points we will add a constant IR
contribution in Sect.~4.4 to arrive at the light curve presented in
Fig.~\ref{FIGURE:bolum}.


     The slope of the late time decline is well fit by a linear decay
in absolute magnitudes by 1.55 magnitudes per 100 days from 140 to 538
days past explosion. This is certainly faster than the decay rate of
the radioactive \element[][56]{Co} powering the supernova at this
phase, again showing that most of the gamma-rays are slipping out of
the ejecta at late phases (S00). We discuss this further in Sect.~5.

\subsection{HST Observations}

     The photometry from the HST observations is presented in
Table~\ref{TABLE:hstphot}.  The CL magnitudes are the AB magnitudes
measured by DAOPHOT on the template subtracted frames and with
aperture corrections applied (CL$_{\mathrm{Sub}}$).  The errors are
simply the formal errors from {\sc DAOPhot} added to the standard
deviations in the aperture corrections.  This may be a bit optimistic,
and from the artificial star test presented in Sect.~3.2 we regard an
additional systematic error of 0.2 magnitudes as a more reliable
estimate of the errors.  In Table~\ref{TABLE:hstphot} we have also
included the magnitudes measured on the drizzled frames using {\sc
DAOPhot} PSF-fitting (CL$_{\mathrm{PSF}}$).  These clearly
overestimate the flux of the supernova by including also light from
the underlying structure, emphasizing the need of template subtraction
even at the resolution achieved by \emph{HST}.


\begin{table}
\begin{center}
\caption{HST/STIS photometry of SN 1998bw.}
\begin{tabular}{lcc}

\hline
\noalign{\smallskip}

Date & CL$_{\mathrm{PSF}}$ & CL$_{\mathrm{Sub}}$ \\

\noalign{\smallskip}
\hline
\noalign{\smallskip}

11 June 00  & $25.33 \pm 0.06$ & $25.90 \pm 0.05$ \\
25 June 00  & $25.45 \pm 0.05$ & $26.03 \pm 0.04$ \\
21 Nov.\ 00 & $26.04 \pm 0.10$ & $26.88 \pm 0.06$ \\
28 Aug.\ 01 & $\sim27.4 \pm 0.1$ & $\ga28.5$ \\

\noalign{\smallskip}
\hline
\end{tabular}
\label{TABLE:hstphot}
\end{center}

AB-magnitudes from the STIS observations. The CLEAR-filter magnitudes
obtained by template subtraction are labelled CL$_{\mathrm{Sub}}$.
The magnitudes estimated directly from PSF-fitting (CL$_{\mathrm{PSF}}$)
clearly overestimate the supernova luminosity.

\end{table}


     The very first result to note is, of course, the clear detection
of the supernova in the subtracted frames (Fig.~\ref{FIGURE:hst}).
Fynbo et~al.\ (2000) used VLT imaging (S00) to establish the position
of the supernova in their STIS images.  Based on this, they suggested
that \object{SN~1998bw} was identical to the object positioned close
to the middle of this messy region.  Our analysis has demonstrated
that this object is indeed fading.
This strongly reinforces the supernova identification and allows us to
follow the supernova to very late phases.

     Secondly, from the magnitudes in Table~\ref{TABLE:hstphot}, and
from Fig.~\ref{FIGURE:lc}, we note that the supernova light curve
seems to be levelling out at the very late phases. The reason for this
will be discussed in Sect.~5.

     In order to test if the supernova was still present in the 2001
Aug.\ 28 CL image we used {\sc DaoPhot II}/{\sc AllStar} to fit a PSF
to the light at the location of \object{SN~1998bw}.  Since there was
no template to subtract from this image, the PSF fit will be strongly
biased by the underlying light at the location of the supernova.  The
aperture-corrected PSF magnitude is CL$_{\mathrm{PSF}} = 27.4 \pm
0.1$.  However, an examination of the residuals after subtracting the
PSF for the supernova suggests that the flux from the supernova is
contaminated by light from the underlying structure and by the object
$0\farcs056$ south of the supernova.  To estimate an upper limit we
again performed an artificial star test by injecting stars of
different magnitudes onto regions of similar background as the
supernova position.  For input magnitudes fainter than $\sim29-30$, we
found that the recovered PSF magnitudes were always close to 27.4, as
measured for the supernova position. Using the standard deviation of
the recovered magnitudes as a one sigma error, we find the $3\sigma$
upper limit of about 28.5 for the supernova. This can be regarded as
an upper limit on the emission from the supernova in the CL band 1221
days past explosion. This suggests that the supernova continued to
fade well into 2001 at a rate similar to that observed between June
and November 2000.

\subsection{Combining the datasets}

     In measuring the ground based photometry, we made the implicit
assumption that the supernova had completely disappeared in our last
VLT template images. From the \emph{HST} results we now know that the
supernova was still there at this phase, although very faint. In fact,
at a magnitude of $\sim26.7$ on day 907 the supernova will not affect
any of the ground based measurements by more than 0.05 magnitudes.  We
will make no correction for this effect.

     However, our late VLT upper limit may indeed be significantly
affected, by up to 0.25 magnitudes. We have taken this into account in
the combined upper limit on the luminosity presented in
Fig.~\ref{FIGURE:bolum}.

     Better spatial resolution clearly allows more accurate
subtractions and thus photometry.  It is clear, however, that even
with \emph{HST} there may be unresolved substructures contributing to
the PSF\@.  At 35 Mpc each drizzled pixel corresponds to 4.3 parsecs
on the sky.  In Table~\ref{TABLE:hstphot} we have included the
magnitudes obtained with PSF-fitting on the drizzled frames, and it is
clear that such an approach overestimates the flux in the supernova.
The effect increases at later phases and would again make the light
curve flatten.  This effect should, however, be correctly accounted
for using the template subtraction. We therefore believe that the
flattening of the late light curve as observed by \emph{HST} is a
robust result.

     There is again the issue of contamination from remaining
supernova flux in the \emph{HST} template frame. Our estimate of the
upper limit on the supernova CL magnitude in the 2001 August frame is
$M^{\mathrm{AB}}_{\mathrm{CL}} \ga 28.5$.
If the supernova was really at this magnitude in the template frame,
the supernova would be brighter than estimated in June and November
2000. The effect could amount to 0.1 magnitudes in June and some
0.2--0.3 magnitudes in November. This is an uncertainty we can not
overcome without further deep and very late imaging. Here we only note
that such a contamination will not dramatically affect the late light
curve slope. It would only flatten the late light curve between 778
and 940 days from a slope of 0.6 magnitudes per 100 days to 0.5
magnitudes per 100 days.

     It is also non-trivial to compare directly the photometry obtained in
the very broad CL filter with the ground based data. Established
conversions as those presented in Rejkuba et~al.\ (2000) are based on
stars, and the late supernova spectrum differs from any stellar
spectrum.  Simply integrating a flat spectrum over the FWHM is a crude
approach for such a wide filter.  Nakamura et~al.\ (2001a) assumed
that ($M_{\mathrm{AB}} =$) $M_{\mathrm{V}} = M_{\mathrm{bol}}$ to put
the photometry of Fynbo et~al.\ (2000) onto their ``bolometric'' light
curve. Such an estimate differs by a factor 1.8 from the simple
integration.

     The correct approach would of course be to use the supernova
spectra obtained at the same epochs for the corrections, but no such
spectra are available. The latest spectrum from S00 is from 504 days
past explosion.  However, this spectrum is rather noisy and may be
contaminated by the nearby objects revealed by \object{HST}.

     We have taken the following approach to compare the different
observations.  From the broad band VLT photometry we do get a gross
spectral energy distribution (SED) for the epochs up to 540 days past
explosion.  This distribution is seen to evolve fairly slowly with
time and should not be influenced by the nearby objects.  At all
epochs, most of the energy (${\nu}F_{\nu}$) emerges in the $R$-band.
From the late-time spectroscopy in S00 we know that this is mainly due
to the strong [\ion{O}{i}] $\lambda\lambda$ 6300,6364 lines.  We will
assume that the same gross spectral distribution is appropriate at the
late \object{HST} phases, with a fairly small contribution from the
$U$-band (P01) and a continuation of the SED into the IR as observed
on day 370.

     We then used {\sc SYNPHOT} to scale this SED to give the measured
count rates for the CL filter function. The scaled SED can then be
integrated, after a correction for extinction (Fitzpatrick 1999), and
converted to luminosity as above.  We noted that the count rate in the
CL filter does not depend much on the assumed SED outside the
$L_{BVRI}$ range.  Assuming no flux outside 4000--8000 {\AA} decreased
the measured counts by only $\sim14$\%, as both the SED and the CL
filter curve peak in the middle of this interval.  To avoid
assumptions on the non-observed SED regimes we therefore integrated
the SED only over the 4000--8000 {\AA} range, to make it directly
comparable to the ground based data.

\subsection{The OIR Light Curve}

     An uncertainty in the bolometric light curve of P01 was the
contribution of the IR emission. They assumed that the fraction of the
$UVOIR$ emission escaping in the IR was the same at late times as it
was on day 65, the last day covered by their IR observations. Then the
IR ($JHK$) contained $\sim35$\% of the supernova emission, and this
constant fraction was added to the optical light curve at all later
phases (P01).

     It is true that for the well observed \object{SN~1987A}, the
fraction of the energy emerging from the supernova in the $V$, $R$,
and $I$ bands ($L_{VRI}$) was virtually constant ($\sim40$\%) up to
400 days after the plateau phase (Schmidt et~al.\ 1993).  However,
the physics of a rapidly expanding, hydrogen-free ejecta will differ
from the case of \object{SN~1987A}.  As the radioactive heating
decreases and the ejecta expand, the gas will cool down and a larger
fraction of the emission may therefore be pushed into the IR\@.  The
assumption of constant IR contribution at late phases should therefore
be checked.

     To estimate the contribution in the IR we followed the same
procedure as outlined above for $L_{BVRI}$. The optical emission was
interpolated on day 370 and integrated from $B$ to $I$. The total
emission was then calculated by including also the $J$ and $H$-bands
in the integral, where the conversions to flux from Wilson et~al.\
(1972) were used.  The result is that at this epoch 42\% of the
emission emerges in the near-IR\@.  The same exercise for the data of
day 65 gives a fraction of 31\% (comparing $L_{BVRI}$ to
$L_{BVRIJH}$).  At all epochs, most of the energy emerges in the
$R$-band.

     It is thus clear that the importance of the near-IR has increased
with time, although the effect is not dramatic.

     To complete our OIR (optical-infrared) light curve we have simply
added the IR fraction observed at day 370 to all the dates on the
light curve presented in Fig.~\ref{FIGURE:lc}. This assumes that the
IR contribution does not evolve at the last \emph{HST} phases.

\section{Discussion}

     The final OIR light curve is admittedly and necessarily based on
a number of assumptions, but we are confident that the basic features
of the light curve, the steep decline followed by a late flattening,
are real.  In this section we will discuss a few possible
interpretations for such an evolution.

     We first note the steep exponential decline seen in the ground
based observations.  This fast decline appears to have started already
about 65 days past the explosion (McKenzie \& Schaefer 1999; S00), and
the light curve continues to fall significantly steeper than the decay
rate of \element[][56]{Co} up to more than 500 days past explosion.
There is thus no sign for a ``positron phase'' in which the fully
deposited kinetic energy from the positrons would dominate the light
curve.  Two different scenarios could account for this.

     One is that the optical depth for gamma-rays, albeit decreasing,
may always be large enough to dominate over the positrons. An example
is shown below (Fig.~\ref{FIGURE:filters}).
Alternatively, the positrons may indeed dominate at late phases, but
some of the positrons are able to escape the ejecta before being
thermalized.

     The second characteristic to note in the very late-time light
curve of \object{SN~1998bw} is the flattening of the light curve at
the last phases, as observed by \emph{HST}. The late-time light curves
of core-collapse supernovae can be powered by several mechanisms (see
e.g., Sollerman et al. 2001), as will be further
discussed below.

\subsection{Radioactive Powering?}

     The contributions from different radioactive elements to late
supernova light curves were discussed by Woosley
et al.
(1989). Only in \object{SN~1987A} has a radioactive decay other than
(\element[][56]{Ni}$\rightarrow$)\element[][56]{Co}$\rightarrow$\element[][56]{Fe}
been unambiguously observed to power the late light curve.  Also for
this supernova, a flattening of the light curve was observed after
about 800 days.  The reason for this was actually two-fold. First, at
these late phases the radioactive decay of the more long-lived nucleus
\element[][57]{Co} became important. Secondly,
the long recombination scale allowed energy stored at earlier epochs
to contribute at later phases (Clayton et~al.\ 1992; Fransson \& Kozma
1993).  This is known as the ``freeze-out'' effect.

     For \object{SN~1998bw}, with its very high explosion energy, the
discussion must also take into account that the optical depth to
gamma-rays decrease very rapidly. The high explosion energy also opens
the possibility for different nucleosynthesis (Nakamura et~al.\
2001b).

\begin{figure*}
\includegraphics{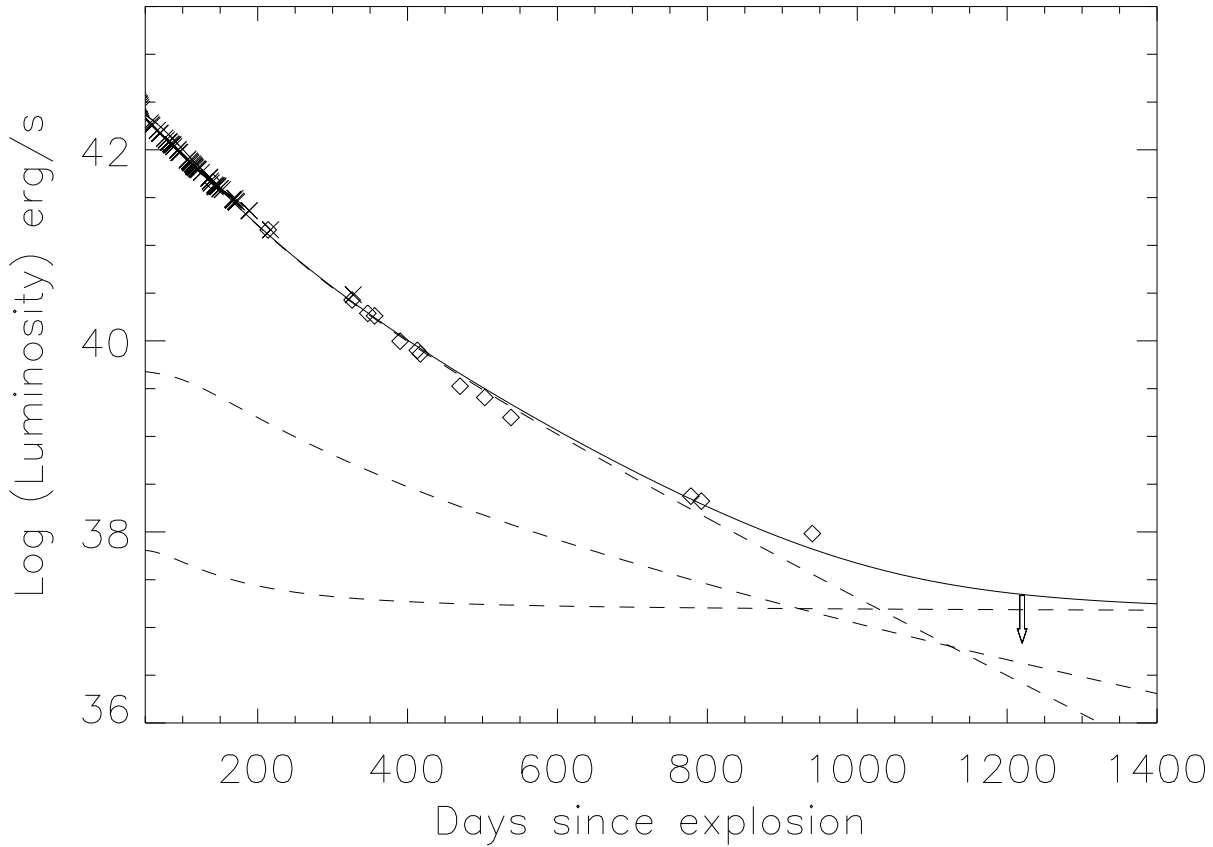}
\caption{Using a very simple model for the radioactive powering of
\protect\object{SN~1998bw}, a reasonable fit to the data can be
achieved. The model is described in the text. The powering of
\element[][56]{Co}, \element[][57]{Co}, and \element[][44]{Ti}
contributes at progressively later phases. The early observations
marked by crosses are from P01 and shifted to a distance of 35 Mpc.  }
\label{FIGURE:filters}
\end{figure*}

     The decreasing optical depth to gamma-rays is obvious from the
first part of the light curve, as mentioned above.  A very simplistic
model to account for this in terms of radioactive decay is presented
in Fig.~\ref{FIGURE:filters}.  In this model the flux from the decay
of \element[][56]{Co} evolves as
$\mathrm{e}^{-t/111.3} \times \left(1 -
0.965\mathrm{e}^{-\tau}\right)$, where the optical depth, $\tau =
{(t_1/t)}^2$, decreases due to the homologous expansion. Here, 111.3
days is the decay time of \element[][56]{Co} and $t_1$ sets the time
when the optical depth to gamma-rays is unity. Furthermore, 3.5\% of
the energy in these decays is in the form of the kinetic energy of the
positrons, which are assumed to be fully trapped.  This model can
reasonably well mimic the observed light curve from day 64 to day 538
for a value of $t_1 = 105$ days (Fig.~\ref{FIGURE:filters}).  This means
that the trapping of gamma-rays is almost complete at day 64, but that
the positron contribution is not dominating until after $\sim 600$
days.


     At the \emph{HST} phases the light curve flattens out and can no
longer be explained in terms of \element[][56]{Co}. An interesting
possibility is then a contribution from \element[][57]{Co}, in
particular since the decay rate in the \emph{HST} detections seems to
agree well with the decay time of
\element[][57]{Co}$\rightarrow$\element[][57]{Fe}.
 
     However, the decay of \element[][57]{Co} has no positron channel.
A fairly large amount of \element[][57]{Ni} (which quickly decays to
\element[][57]{Co}) must therefore be synthesized in the explosion, in
order to make a significant contribution in this model.

     In Fig.~\ref{FIGURE:filters} we have included a ratio of
\element[][57]{Ni}/\element[][56]{Ni} which is three times larger than
the ratio observed in \object{SN~1987A}.  The optical depth to gamma
rays is also enhanced by a factor of 2.4 as compared to
\element[][56]{Ni}, to allow for the smaller gamma-ray energies
involved (Woosley et~al.\ 1989). In addition, \element[][44]{Ti} has
also been added with a three times larger ratio. This nucleus
contributes at much later phases in \object{SN~1987A} (Kozma 2000;
Lundqvist et~al.\ 2001), but can become important at relatively
earlier phases in \object{SN~1998bw} as it has a significant positron
channel.

     Although these abundance enhancements are quite arbitrarily
chosen to match the observations, we emphasize that the
nucleosynthesis yields may indeed be different for very energetic and
possibly asymmetric explosions. Nakamura et~al.\ (2001b) and Maeda
et~al.\ (2001) clearly show that enhanced amounts of $^{44}$Ti may be
expelled in such circumstances.  We therefore note the possibility
that the late light curve of \object{SN~1998bw} can be explained in
terms of radioactivity, if the abundances of the radioactive isotopes
are enhanced compared to the case of \object{SN~1987A}.

     Although the above model is clearly too simplistic to give
accurate results, it can be used to estimate the amount of
\element[][56]{Ni} needed to power the supernova at a distance of 35
Mpc. With the radioactive luminosities properly included (Kozma \&
Fransson 1998) we have used a nickel mass of $0.3~{~\rm M}_\odot$ to power
the light curve in Fig.~\ref{FIGURE:filters}.

     This is close to the lower limit obtained by S00.  This is indeed
a lower limit, as the simple model forces the gamma-rays to be almost fully
trapped at day 65, at the start of the rapid decline. The models
presented in S00, with a complete calculation of the gamma-deposition
for real explosion models, trapped 6--10\% of the gamma-ray energy at
400 days, while the toy model presented here traps $\sim7$\% at this
epoch. Although detailed calculations are needed to accurately
determine the nickel mass, it is clear that the estimates will be
lower than in S00 simply because some of the light attributed to the
supernova in that analysis was due to the complex background.

     The above estimate is, however, not very
far from the 0.4 ${~\rm M}_\odot$ required to power the peak of the light
curve in the refined models of Nakamura et~al.\ (2001a), especially
when the different distance estimates are taken into account.

     We note that a scenario where the steep light curve decline is
instead due to positron escape will clearly require much higher nickel
masses to account for the observed luminosity. In such a scenario the
late time contribution of \element[][57]{Co} is less likely, because
with no positron channel an unrealistically high abundance of this
isotope would be required.

\subsection{Other Powering Mechanisms?}

     It must be clarified that apart from radioactive heating, several
other mechanisms could power the late light curve. None of the usual
suspects can really be ruled out based on the sparse observational
constraints we have at the latest phases.

     Interaction with circumstellar material (CSM) is not uncommon in
core-collapse supernovae (e.g., Leibundgut 1994). In the context of
\object{SN~1998bw}, Tan
et al.
(2001) proposed an
association of \object{GRB~980425} based on a model with a high
pre-supernova mass loss of $\sim$few $\times 10^{-4}$
${~\rm M}_\odot$~yr$^{-1}$.  Weiler
et al.
(2001) also argued
for a CSM from radio monitoring of the supernova.  However, the very
fast wind expected from a WR progenitor can give a rather modest
density even with a high mass loss rate, and the interaction need not
give rise to optical emission.  We note that as long as the supernova
was spectroscopically monitored, no spectral signatures from CSM
interaction were ever detected in \object{SN~1998bw} (S00).  Although
absence of evidence is not evidence of absence, this may constrain the
most massive progenitor models.

     If \object{SN~1998bw} formed a black hole which accretes matter,
this could in principle show up in the late light curve (Fynbo et~al.\
2000). We note that the luminosity at late phases is close to the
Eddington luminosity, which would require very high
efficiency. Moreover, an accretion scenario should provide a power-law
decay rate (Balberg
et al.
2000)
, which is not
observed. Black hole powering is mainly expected to be seen for
supernovae with very little radioactive material, because the
radioactive heating will otherwise outshine the accretion
luminosity. This is quite the opposite of \object{SN~1998bw}.

     Another scenario for late-time emission is a light echo, as
observed in \object{SN~1998bu} (Cappellaro et~al.\ 2001). Again, we
have little information available to exclude such a possibility. We
can only note that the supernova appears fairly red in the day 778
data (LP versus CL), while a light echo should reflect an early blue
phase.

     Finally, the late light curve might also be boosted by a
freeze-out effect. As mentioned above, this was observed in
\object{SN~1987A} and contributed to the light curve tail powered by
\element[][57]{Co}, as illustrated in Fig.~\ref{FIGURE:hst} of
Fransson \& Kozma (1993).  Most of the freeze-out in \object{SN~1987A}
took place in the extended hydrogen envelope, which is absent in
\object{SN~1998bw}. On the other hand, the faster expansion of SN
1998bw favors a freeze-out scenario.  If freeze-out was indeed
important at these phases, smaller yields of the long-lived nuclei
would be required to power the light curve. Detailed time-dependent
modeling is needed to quantify this effect.

\section{Summary}

     We have presented new and updated photometry of the famous GRB
supernova \object{SN~1998bw}. Using the template subtraction technique
we have revised the late-time light curve and shown that the fast
decline rate continues to more than 500 days past explosion.  This
highlights the observational problem of accurate photometry in complex
regions, where core-collapse supernovae are known to reside. We note
that no sign of a fully trapped \element[][56]{Co} positron phase is
observed.  A simple radioactive model with $\sim 0.3$ ${~\rm M}_\odot$ of
ejected \element[][56]{Ni} can fit the data, and can be regarded as a
lower limit to the amount of ejected nickel in \object{SN~1998bw}.
Our very late HST observations allowed us to follow the supernova to
$\sim 1000$ days, and to detect a late flattening of the light
curve. One of the possible scenarios to explain this is the
contribution of radioactive isotopes with longer life times. This
could require different nucleosynthesis for the very energetic
\object{SN~1998bw} than for \object{SN~1987A}.

\begin{acknowledgements}
     We thank Paolo Mazzali and Keiichi Maeda for interesting
discussions, and all members of the SINS team for support.  This
research was supported by NASA through grant GO-2563.001 to the SINS
group from STSCI, which is operated by AURA, Inc., under NASA contract
NAS5-26555.  A.V.F. acknowledges the support of the Guggenheim
Foundation Fellowship.  S.T.H. acknowledge support from the NASA LTSA
grant NAGS--9364.
\end{acknowledgements}




\begin{thebibliography}{}

\bibitem[2000]{BZS2000} 
     Balberg, S., Zampieri, L. \& Shapiro, S~L. 2000, \apj, 541, 860

\bibitem[1979]{B1979}
     Bessell, M. S. 1979, \pasp, 91, 589

\bibitem[2001]{CPM2001}
     Cappellaro, E., Patat, F., Mazzali, P.~A., et~al.\ 2001, \apj,
     549, 215 

\bibitem[1992]{CLT1992}
     Clayton, D.~D., Leising, M.~D., The, L.-S., Johnson, W.~N., \&
     Kurfess, J.~D. 1992, \apj, 399, 141

\bibitem[2001]{CSP2001}
     Clocchiatti, A., Suntzeff, N.~B., Phillips, M.~M., et~al.\ 2001,
     \apj, 553, 886

\bibitem[1986]{FPS1986}
     Filippenko, A.~V., Porter, A.~C., Sargent, W.~L.~W., \&
     Schneider, D.~P. 1986, \aj, 92, 1341

\bibitem[1999]{F1999}
     Fitzpatrick, E.~L. 1999, \pasp, 111, 63

\bibitem[1993]{F1993}
     Fransson, C. \& Kozma, C. 1993, \apjl, 408, L25

\bibitem[2002]{FH2002}
     Fruchter, A.~S., \& Hook, R.~N. 2002, \pasp, 114, 144

\bibitem[2000]{FHA2000}
     Fynbo, J.~U., Holland, S.~T., Andersen, M.~I. et~al.\ 2000,
     \apjl, 542, L89 

\bibitem[1998a]{GVP1998}
     Galama, T.~J., Vreeswijk, P.~M., Pian, E., et~al.\ 1998a,
     \iaucirc 6895

\bibitem[1998b]{GVV1998}
     Galama, T.~J., Vreeswijk, P.~M., van Paradijs, J., et~al.\ 1998b,
     \nat, 395, 670 

\bibitem[2000]{GBB200}
     Gardner, J.~P., Baum, S.~A., Brown, T.~M., et~al.\ 2000, \aj,
     119, 486

\bibitem[2000]{HFT2000}
     Holland, S.~T, Fynbo, J.~U., Thomsen, B., et~al.\ 2000, GCN Circ.\
     698 

\bibitem[1999]{HWW1999}
     H\"oflich, P., Wheeler, J.~C., \& Wang, L. 1999, \apj, 521, 179


\bibitem[1998]{IMN1998}
     Iwamoto, K., Mazzali, P.~A., Nomoto, K., et~al.\ 1998, \nat, 395,
     672 

\bibitem[2000]{K2000}  
     Kozma, C. 2000, in Future Directions of Supernova Research:
     Progenitors to Remnants, eds. S. Cassisi \& P. Mazzali, in
     Memorie della Societ{\`a} Astronomica Italiana

\bibitem[1998]{KF1998}
     Kozma, C., \& Fransson, C. 1998, \apj, 496, 946

\bibitem[1998]{KFW1998}
     Kulkarni, S.~R., Frail, D.~A., Wieringa, M.~H., et~al.\ 1998,
     \nat, 395, 663 

\bibitem[1994]{L1994}
     Leibundgut, B. 1994, in Circumstellar Media in the Late Stages of
     Stellar Evolution, eds. R.~E.~S. Clegg, I.~R. Stevens, \&
     W.~P.~S. Meikle (Cambridge: Cambridge Univ.\ Press), 100

\bibitem[2001]{LKS2001}
     Lundqvist, P., Kozma, C., Sollerman, J., \& Fransson, C. 2001,
     \aap, 374, 629

\bibitem[2002]{MNN2002}
     Maeda, K., Nakamura, T., Nomoto, K., Mazzali, P.~A., Patat, F.,
     \& Hachisu, I. 2002, \apj, 565, 405

\bibitem[2001]{MNP2001}
     Mazzali, P.~A., Nomoto, K., Patat, F., \& Maeda, K. 2001, \apj,
     559, 1047

\bibitem[1999]{MS1999}
     McKenzie, E.~H. \& Schaefer, B.~E. 1999, \pasp, 111, 964

\bibitem[2001a]{NaM2001}
     Nakamura, T., Mazzali, P.~A., Nomoto, K., \& Iwamoto, K. 2001a,
     \apj, 550, 991

\bibitem[2001b]{NUI2001}
     Nakamura, T., Umeda, H., Iwamoto, K., Nomoto, K., Hashimoto, M.,
     Hix, W.~R., \& Thielemann, F.-K.  2001b, \apj, 555, 880 

\bibitem[2001]{NoM2001}
     Nomoto, K., Mazzali, P.~A., Nakamura, T., Iwamoto, K., Danziger,
     I.~J., \& Patat, F. 2001, in STScI Symp.\ Ser.\ 13, Supernovae \&
     Gamma-Ray Bursts, eds. M. Livio, N. Panagia, \& K. Sahu
     (Cambridge: Cambridge Univ.\ Press)

\bibitem[2001]{PCD2001}
     Patat F., Cappellaro, E., Danziger, J., et~al.\ 2001, \apj, 555,
     900 

\bibitem[2000]{RMG2000}
     Rejkuba, M., Minniti, D., Gregg, M.~D., et~al.\
     2000, \aj, 120, 801

\bibitem[1993]{SMS1993}
     Schechter, P.~L., Mateo, M., \& Saha, A. 1993, \pasp, 105, 1342

\bibitem[1998]{SFD1998}

     Schlegel, D.~J., Finkbeiner, D.~P., \& Davis, M. 1998, \apj, 500,
     525

\bibitem[1998]{SSP1998}
     Schmidt, B.~P., Suntzeff, N.~B., Phillips, M.~M., et~al.\ 1998,
     \apj, 507, 46 

\bibitem[1993]{SKS1993}
     Schmidt, B.~P., Kirshner, R.~P., Schild, R., et~al.\ 1993, \aj,
     105, 2236 

\bibitem[1998]{SFP1998}
     Soffitta, P., Feroci, M., Piro, L., et~al.\ 1998, \iaucirc 6884



\bibitem[2000]{SKF2000}
     Sollerman, J., Kozma, C., Fransson, C., et~al.\ 2000, \apj, 537,
     127 

\bibitem[2001]{SKL2001}
     Sollerman, J., Kozma, C., \& Lundqvist, P. 2001, \aap, 366, 197

\bibitem[2002]{S2002}
     Sollerman, J. 2002, New Astronomy Reviews, March issue
 
\bibitem[1987]{S1987}
     Stetson, P.~B. 1987, \pasp, 99, 191

\bibitem[1988]{SH1988} 
     Stetson, P. B. \& Harris, W. E., 1988, \aj, 96, 909\\

\bibitem[2001]{TMM2001}
     Tan, J.~C., Matzner, C.~D., \& McKee, C.~F. 2001, \apj, 551, 946


\bibitem[2001]{WPM2001}
     Weiler, K.~W., Panagia, N., \& Montes, M.~J. 2001, \apj, 562,
     670

\bibitem[1999]{WES1999}
     Woosley, S.~E., Eastman, R.~G., \& Schmidt, B.~P. 1999, \apj,
     516, 788

\bibitem[1989]{WHP1989}
     Woosley, S.~E., Hartmann, D., \& Pinto, P.~A. 1989, \apj, 346,
     395

\bibitem[1972]{WSN1972}

     Wilson, W.~J., Schwartz, P.~R., Neugebauer, G., Harvey, P.~M., \&
     Becklin, E.~E. 1972, \apj, 177, 523

\end{thebibliography}
\end{document}